%% file: sample-sigconf.tex
\documentclass[sigconf]{acmart}

\AtBeginDocument{%
  }

\copyrightyear{2025} 
\acmYear{2025} 
\setcopyright{cc}
\setcctype{by-nc-nd}
\acmConference[ICMI Companion '25]{Companion Proceedings of the 27th International Conference on Multimodal Interaction}{October 13--17, 2025}{Canberra, ACT, Australia}
\acmBooktitle{Companion Proceedings of the 27th International Conference on Multimodal Interaction (ICMI Companion '25), October 13--17, 2025, Canberra, ACT, Australia}
\acmDOI{10.1145/3747327.3764897}
\acmISBN{979-8-4007-2076-5/2025/10}

\begin{document}

\title{MultiGen: Child-Friendly Multilingual Speech Generator with LLMs}

\author{Xiaoxue Gao, Huayun Zhang and Nancy F. Chen}
\email{Gao\textunderscore Xiaoxue@i2r.a-star.edu.sg, Zhang\textunderscore Huayun@i2r.a-star.edu.sg, nfychen@i2r.a-star.edu.sg}
\affiliation{%
  \institution{Institute for Infocomm Research, Agency for Science, Technology, and Research (A*STAR), Singapore}\country{}
}

\begin{abstract}
Generative speech models have demonstrated significant potential in improving human-machine interactions, offering valuable real-world applications such as language learning for children. However, achieving high-quality, child-friendly speech generation remains challenging, particularly for low-resource languages across diverse languages and cultural contexts. 
In this paper, we propose \textit{MultiGen}, a multilingual speech generation model with child-friendly interaction, leveraging LLM architecture for speech generation tailored for low-resource languages. 
We propose to integrate age-appropriate multilingual speech generation using LLM architectures, which can be used to facilitate young children's communication with AI systems through culturally relevant context in three low-resource languages: Singaporean accent Mandarin, Malay, and Tamil. Experimental results from both objective metrics and subjective evaluations demonstrate the superior performance of the proposed \textit{MultiGen} compared to baseline methods.

\end{abstract}



\maketitle

\section{Introduction}
Generative speech models have demonstrated significant potential in improving human-machine interactions across a wide range of applications. Text-to-speech (TTS) systems are increasingly deployed in virtual assistants, accessibility technologies for voice-enabled applications, and interactive AI companions. The growing demand for natural and engaging human-AI communication has driven research toward more natural and personalized speech generation models that can adapt to diverse user demographics and linguistic contexts. For instance, when building AI-based language learning tools \cite{liu2025singakids}, generative speech models \cite{nose2007style,gao2024emo,yasuda2023text,gao2025ttslow,um2020emotional,li2024mm,wang2018style} play a critical role in facilitating language acquisition, enabling children to engage effectively in listening and speaking activities across multiple languages and cultural contexts \cite{zhang2021ai,pine2025speech,chen2024voicebench}.

\begin{figure*}[t]
\vspace{-0.5cm}
\centering
\includegraphics[width=154mm]{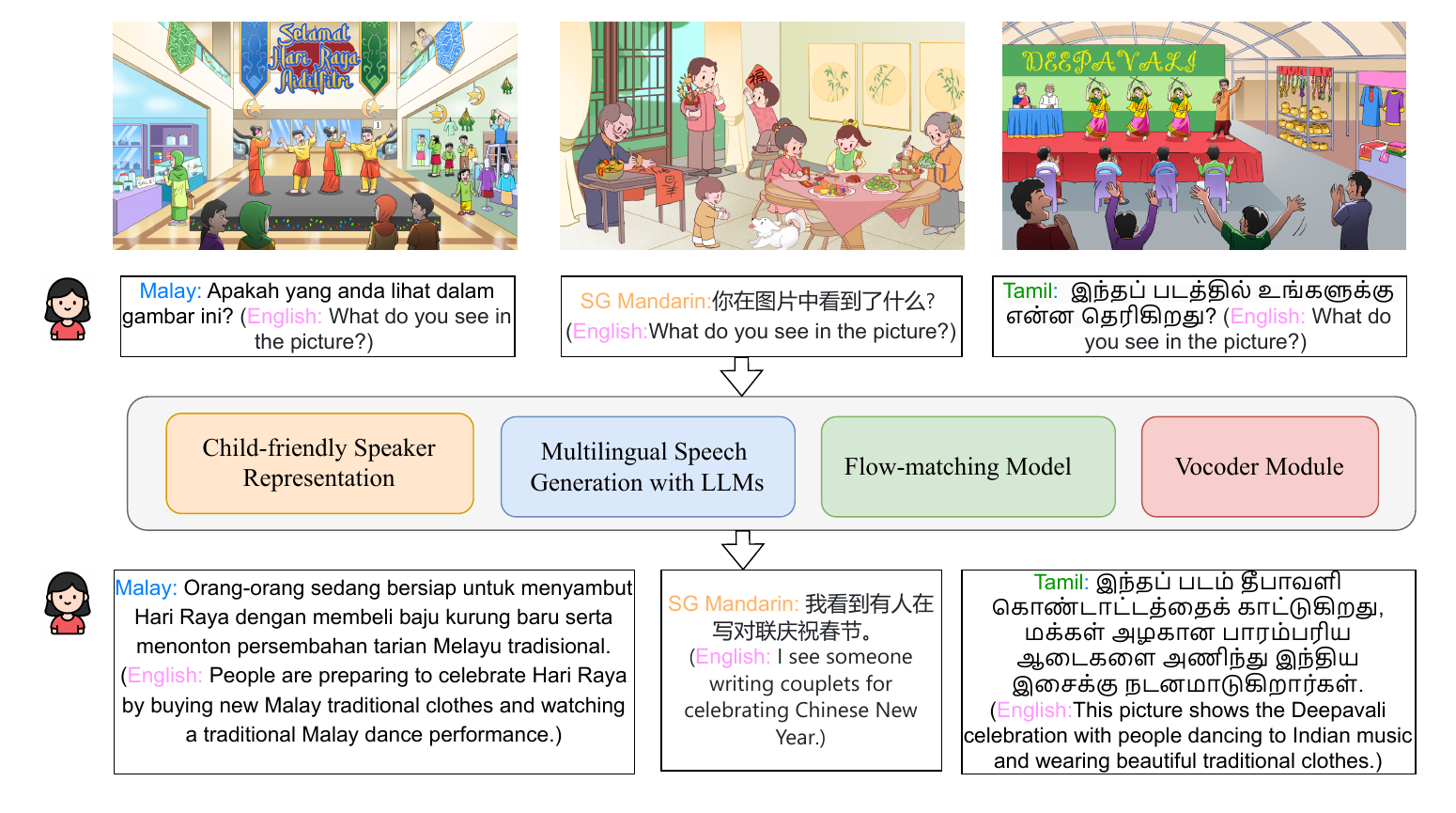}
\vspace{-0.66cm}
\caption{The overview of the proposed speech generation model \textit{MultiGen} for Singaporean-accented Mandarin, Malay and Tamil in multicultural contexts. Transcribed English text is provided for all three languages to aid comprehension.}
\label{overall}
\end{figure*}

Despite advances in adult speech synthesis \cite{lei2021fine,chen2023vector,liu2024emotion,chen2024vall}, child-specific speech data remains scarce \cite{zhang2025automated}, making children one of the most underrepresented groups in speech generation. This scarcity arises primarily from ethical and practical challenges associated with collecting and annotating children's speech \cite{zhang2025automated,kulkarni2025children}. Furthermore, children's speech exhibits unique acoustic and linguistic characteristics such as higher pitch, smaller vocal tracts, and evolving pronunciation patterns distinct from adult speech \cite{zhang2025automated,kulkarni2025children}. Generative speech models often lack adequate child-focused training data, an issue particularly acute in low-resource languages \cite{gong2024initial}.

Multilingual communication is vital in culturally diverse societies like Singapore, where the national curriculum expects children to learn multiple languages—English, Singaporean-accented Mandarin, Malay, and Tamil—to preserve heritage and foster cross-cultural communication \cite{huzaifah2024towards,huzaifah2024meralion,he2024meralion,lovenia2024seacrowd,zhang2021multilingual}. While English dominates daily use, the other three are low-resource in terms of digital tools and learning materials and remain crucial for academic assessment and cultural inclusion \cite{huzaifah2024towards,huzaifah2024meralion,he2024meralion,lovenia2024seacrowd}. 
However, it is challenging for speech generation models in consistent and effective multilingual language learning, especially for under-resourced languages in child-centric contexts \cite{gong2024initial}. The only prior work on speech generation for low-resource languages in educational contexts relies on conventional architectures such as FastSpeech \cite{ren2019fastspeech} and FastPitch \cite{lancucki2021fastpitch}, targeting languages like Onkwawenna Kentyohkwa and Kanyen’kéha in the UK and US \cite{kulkarni2025children}. In contrast, speech generation for Southeast Asian low-resource languages—with 671 million people for 8.75\% of the global population \cite{lovenia2024seacrowd}—remains largely underexplored, particularly with the use of state-of-the-art large language model (LLM) architectures over conventional neural networks.

To bridge this gap, there is a pressing need for a child-friendly, multilingual speech generator in low-resource languages and culturally aware manner. In this paper, we propose \textit{MultiGen}, a multilingual speech generator powered by LLMs, which is motivated from supporting culturally grounded learning experiences, age-appropriate language learning for children \cite{liu2025singakids}. Our \textit{MultiGen} integrates child-friendly, age-appropriate speech generation using an LLM-based neural architecture, leveraging LLM’s in-context learning to flexibly adapt to different language inputs.

The main contributions of this paper are: (1) we propose a multilingual speech generator leveraging an LLM-based architecture; (2) we introduce an age-appropriate training strategy that advances child-friendly speech generation, particularly for low-resource languages; and (3) extensive experiments show that the proposed \textit{MultiGen} significantly outperforms baseline text-to-speech models.

\section{Methodology: \textit{MultiGen}}

We propose \textit{MultiGen}, a child-friendly multilingual speech generator, particularly improving the naturalness and quality on low-resource languages. In practice, when applying such improved text-to-speech models, it enables young children to engage in age-appropriate speaking and listening activities across Tamil, Singaporean-accented Mandarin, and Malay, aligning closely with Singapore's diverse linguistic and cultural landscape \cite{liu2025singakids}. Figure~\ref{overall} illustrates the overall architecture of the \textit{MultiGen} approach in a multicultural setting.

\subsection{Overview}
As illustrated in Figure~\ref{overall}, \textit{MultiGen} integrates low-resource language speech generators with child-friendly voice. MultiGen enables the generation of "What do you see in the picture?" to be spoken in three low-resource languages for children, encouraging children’s listening, oral expression, vocabulary building, and confidence in multilingual communication. The ideal student response, also shown in Figure~\ref{overall}, can likewise be generated by MultiGen to provide children with the correct answers across three languages.

Our \textit{MultiGen} is designed to facilitate multilingual text-to-speech generation, converting input text into child-friendly speech across three target languages. The approach integrates multilingual speech token generation models with LLMs, a flow-matching model, and a vocoder module. The high speech quality can support educational applications, students can practice verbal responses and listen to the generated references in all three languages, enabling them to refine and correct their replies.

\subsection{Child-friendly LLM-based Speech Generator}
Motivated by the successful integration of LLMs in emotional speech synthesis \cite{gao2024emo,gao2025prompt}, we propose an innovative, child-friendly, and age-appropriate training strategy for multilingual speech generation. Our method introduces LLM architectures for the first time targeted at low-resource languages: Malay, Tamil, and Singapore-accented Mandarin. 

Unlike traditional speech generation models like Fastspeech \cite{kulkarni2025children,ren2019fastspeech}, our \textit{MultiGen} integrates LLM and captures aspects of children's speech, such as higher pitch, distinctive prosodic patterns, and intonation, in a data-driven manner. Leveraging the autoregressive nature of LLMs, our \textit{MultiGen} generates child-like multilingual speech outputs by minimizing the Kullback–Leibler (KL) divergence between the predicted and ground-truth probability distributions of multilingual speech tokens, guided by three language identifiers: Tamil, Malay, and Singaporean-accented Mandarin. This KL-based objective encourages the model to approximate natural children's speech more closely. The child-friendly mechanism is achieved by incorporating speaker representations derived from child voices, where x-vector embeddings are extracted from Malay, Tamil, and Singaporean-accented children’s speech to capture age-specific vocal characteristics.

The better text-to-speech quality can foster an immersive and culturally grounded communication experience. To illustrate this advantage, Figure~\ref{overall} presents some representative outputs depicting culturally significant traditional festivals for each linguistic group: Deepavali for the Indian community, Chinese New Year for the Chinese community, and Hari Raya for the Malay community, where students can hear culturally relevant responses—such as buying Malay clothes for Hari Raya, writing couplets for Chinese New Year, and dancing to Indian music for Deepavali.
In practice, we adopt and assess it \textit{MultiGen} on our established work on language learning \cite{liu2025singakids}, which fosters an engaging environment that eases cognitive load and builds children's confidence and language skills through child-to-child communication, demonstrating strong potential for real-world applications in elementary schools across Southeast Asia.

\begin{figure}[t]
\vspace{-0.4cm}
\centering
\includegraphics[width=74mm]{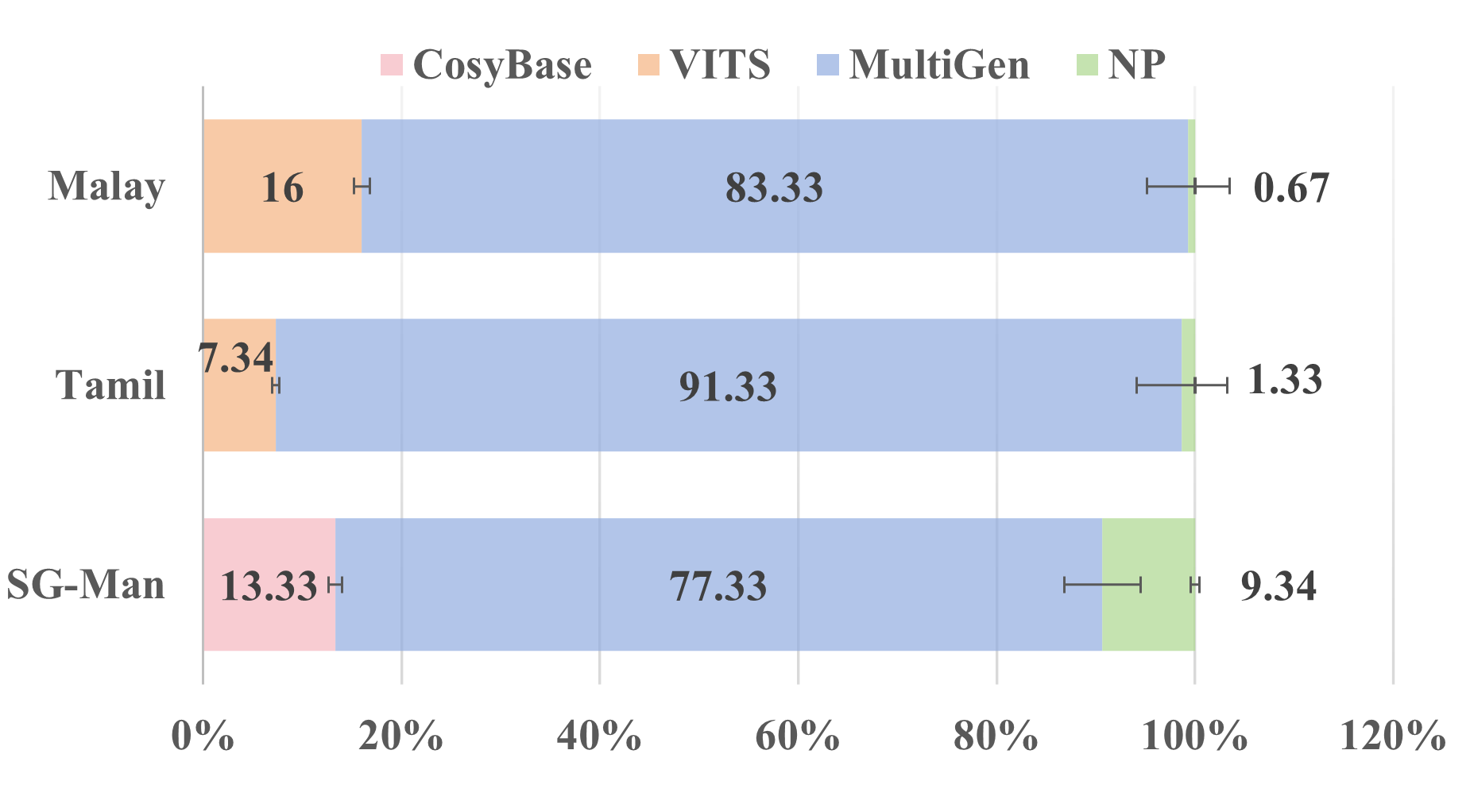}
\vspace{-0.3cm}
\caption{Comparison of AB Preference Tests with 95\% Confidence Intervals: (1) VITS vs. \textit{MultiGen} for Malay; (2) VITS vs. \textit{MultiGen} for Tamil; and (3) CosyBase vs. \textit{MultiGen} for Singaporean-accented Mandarin (SG-Man). ‘NP’ denotes no preference.}
\label{AB}
\vspace{-0.4cm}
\end{figure}

\section{Experiments}
\vspace{-0.1cm}
In this section, we describe the experimental datasets, baselines and experimental setups.
\vspace{-0.3cm}
\subsection{Datasets and Baselines}
We use two state-of-the-art models as baselines: VITS \cite{kim2021conditional} for Malay and Tamil, and CosyBase (CosyVoice-300M) \cite{du2024cosyvoice} for Singaporean accent Mandarin. Notably, in our established language learning system SingaKids \cite{liu2025singakids}, we adopted the VITS model considering both computational efficiency and text-to-speech quality. Moreover, CosyBase does not support Malay or Tamil, while SingaKids does not cover Singaporean accented Mandarin for speech generation \cite{du2024cosyvoice,liu2025singakids}.
We collected, annotated, and curated multilingual text and speech data from children and adults in three languages. The Malay children’s corpus consists of 30,000 utterances from 104 Malay speakers aged 9–16. The Malay adult data comprises 23,897 utterances from 7 speakers, while the Tamil adult corpus includes 33,830 utterances from a single speaker.
The Singaporean-accented Mandarin data contains 1,400 utterances from one child speaker.
\vspace{-0.2cm}
\subsection{Experimental Setups}
The proposed \textit{MultiGen} integrates transformer-based LLM architectures, a pretrained flow-matching model, and a pretrained HiFi-GAN vocoder \cite{kong2020hifi} for multilingual speech generation. The proposed \textit{MultiGen} is fine-tuned from the CosyVoice-300M \cite{du2024cosyvoice} for five epochs using dynamic batching, with the best-performing checkpoint for each language selected based on validation performance. Language identifiers are set to zh, ma, and ta to represent Singaporean-accented Mandarin, Malay, and Tamil, respectively. We conduct extensive listening tests, including AB preference tests~\cite{gao2020personalized}, mean opinion score (MOS) tests~\cite{gao2019speaker}, and speech intelligibility evaluations. The tests involved 30 listeners—10 native Malay, 10 native Tamil, and 10 native Singaporean Chinese speakers—who each assessed 30 child speech samples in their respective languages.

\begin{figure}[t]
\vspace{-0.4cm}
\centering
\includegraphics[width=70mm]{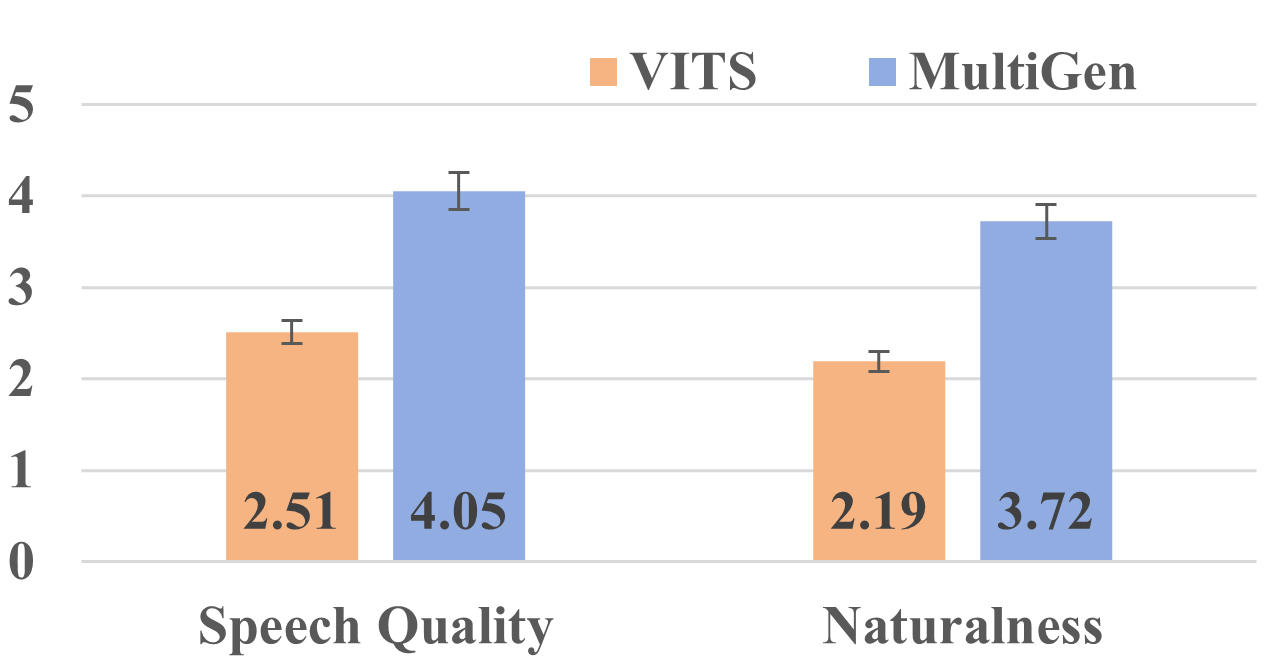}
\vspace{-0.3cm}
\caption{Comparison of speech quality and naturalness results  on MOS with 95\% Confidence Intervals between VITS and \textit{MultiGen} for Tamil. }
\label{mos_ta}
\vspace{-0.2cm}
\end{figure}
\vspace{-0.2cm}
\section{Results and Discussion}
We examine the impact of multilingual human preference on different approaches. We also assess speech quality and speech naturalness using MOS, and analyze speech intelligence through both subjective evaluations and objective Character Error Rate (CER) measurements across multilingual settings. 
\vspace{-0.1cm}
\subsection{Multilingual Human Preference Evaluation}
To evaluate the effectiveness of \textit{MultiGen}, we conduct multilingual human preference evaluations using AB preference tests, where native speakers of Singaporean-accented Mandarin, Malay, and Tamil choose the better speech sample between our model and a baseline (CosyBase or VITS).
As shown in Figure~\ref{AB}, 83.33 \% of listeners choose \textit{MultiGen} over the VITS baseline for Malay, while only 16.00 \% opt for the latter. Similarly, 77.33\% of native human listeners select \textit{MultiGen} over CosyBase for Singaporean accented Mandarin, with only 13.33 \% selecting the baseline. For Tamil, 91.33 \% of participants select \textit{MultiGen} compared to just 7.34 \% for VITS. These findings indicate that \textit{MultiGen} consistently outperforms baseline models across linguistically and culturally diverse settings, highlighting the effectiveness of our age-appropriate, multilingual speech generation design powered by LLMs.

\vspace{-0.1cm}
\subsection{Speech Quality and Naturalness}

To evaluate speech quality and naturalness, we conduct separate MOS tests across the three target languages, asking listeners to rate the samples from 1 (bad) to 5 (excellent).
\vspace{-0.1cm}
\subsubsection{Generative Speech Evaluation for Tamil Language}

Figure~\ref{mos_ta} presents the evaluation results for generated Tamil speech. The proposed \textit{MultiGen} achieves significantly higher scores in both speech quality (4.05) and naturalness (3.72) compared to the VITS baseline, which scores 2.51 and 2.19, respectively. These results demonstrate the effectiveness of \textit{MultiGen} in capturing Tamil linguistic and cultural characteristics of Tamil through a generative speech model based on LLM neural architectures.
This advancement underscores the model's potential to support culturally aware speech generation and promote research for the low-resource Tamil language within the speech community.

\begin{figure}[t!]
\vspace{-0.4cm}
\centering
\includegraphics[width=73mm]{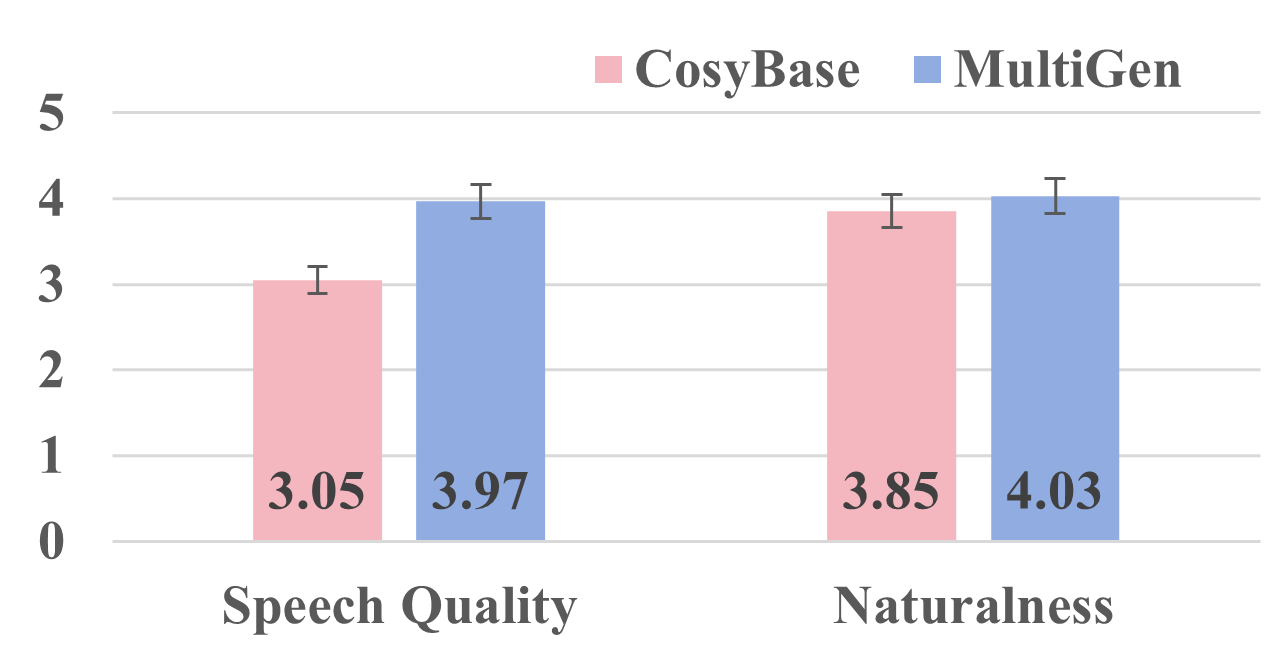}
\vspace{-0.3cm}
\caption{Comparison of speech quality and naturalness results on MOS Scores with 95\% Confidence Intervals between CosyBase and \textit{MultiGen} for Singaporean-accented Mandarin (SG-Man). }
\label{mos_zh}
\vspace{-0.3cm}
\end{figure}
\subsubsection{Generative Speech Evaluation for Singaporean-Accented Mandarin}
Figure~\ref{mos_zh} presents the subjective evaluation results for synthesized Singaporean-accented Mandarin. The proposed \textit{MultiGen} achieves higher scores in both speech quality (3.97 vs. 3.05) and naturalness (4.03 vs. 3.85) compared to the CosyBase baseline. These results demonstrate the effectiveness of our age-appropriate generative speech model adaptation mechanism in producing child-friendly speech suitable in the Singaporean-accented Mandarin context.

\subsubsection{Generative Speech Evaluation for Malay Language}
\begin{figure}[t!]
\vspace{-0.5cm}
\centering
\includegraphics[width=71mm]{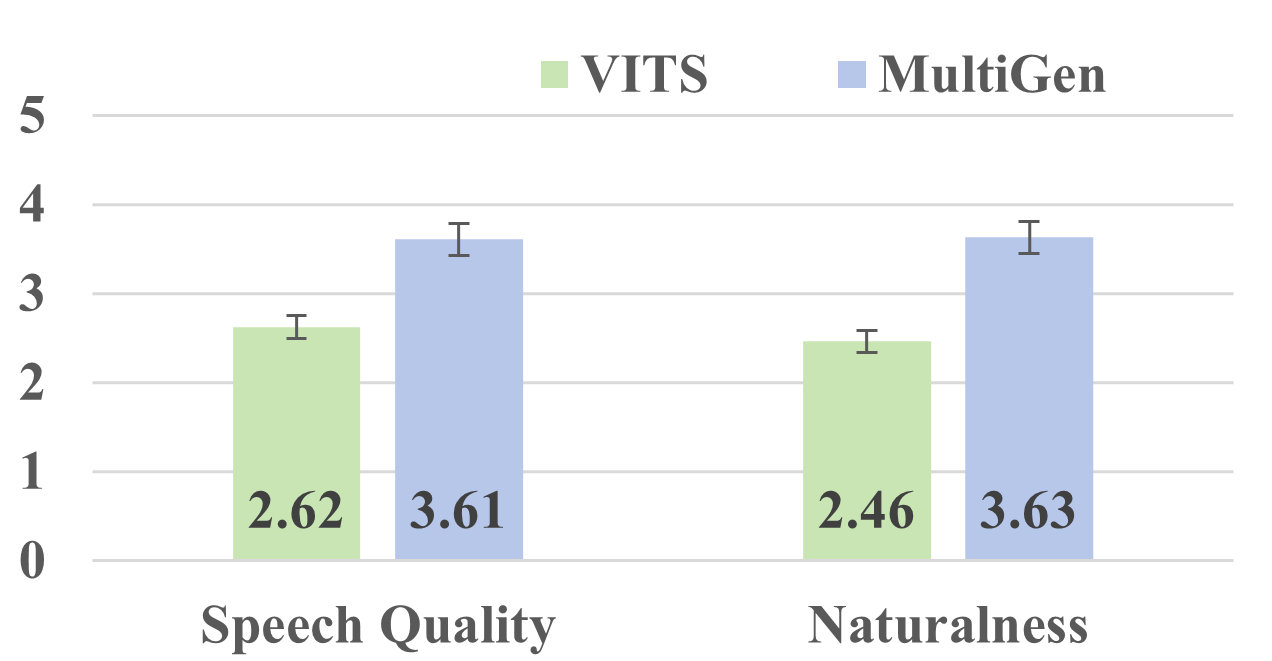}
\vspace{-0.3cm}
\caption{Comparison of Speech Quality and Naturalness Results  Based on MOS Scores with 95\% Confidence Intervals between VITS and the proposed \textit{MultiGen} for Malay. }
\label{mos_ma}
\vspace{-0.2cm}
\end{figure}

Figure~\ref{mos_ma} presents the evaluation results for generative Malay speech. The proposed \textit{MultiGen} achieves higher scores in both speech quality (3.61 vs. 2.62) and naturalness (3.63 vs. 2.46) compared to the VITS.
These findings reflect the effectiveness of \textit{MultiGen} in generating culturally appropriate Malay speech, extending its applicability beyond Tamil and Mandarin to the low-resource Malay language. This advancement underscores the potential of the approach to contribute to culturally inclusive speech generation research for underrepresented languages.

\vspace{-0.1cm}
\subsection{Multilingual Speech Intelligence}
To comprehensively assess the pronunciation accuracy of \textit{MultiGen}, we perform both objective and subjective evaluations across the three low-resource languages, as presented in Table~\ref{cer}. For objective evaluations, we utilize automatic speech recognition (ASR) for each language: the Tamil and Malay ASR models we developed in SingaKids \cite{liu2025singakids}, and the Whisper model\footnote{\url{https://huggingface.co/openai/whisper-large-v3}} for Mandarin. The synthesized speech samples are transcribed using these models, and CER is calculated by comparing the transcriptions to the ground-truth text. A lower CER reflects higher pronunciation accuracy.

However, in low-resource scenarios, ASR systems often exhibit limited robustness and may not fully capture the nuances of pronunciation accuracy, particularly in children's speech. To address these limitations, we complement the objective evaluation with human assessments. For the subjective evaluation, native speakers are asked to rate the pronunciation accuracy—referred to as human speech intelligence—on a scale from 0 (completely inaccurate) to 100 (fully accurate). These human ratings offer more culturally and linguistically grounded insights, reflecting alignment with native pronunciation norms across the respective languages.

Table~\ref{cer} shows that the proposed \textit{MultiGen} achieves consistently better performance than the baselines across all three languages, as measured by both objective evaluations using CER and subjective evaluations by human listeners. In terms of CER, \textit{MultiGen} substantially reduces recognition errors compared to the baselines across three languages. These improvements reflect a higher degree of pronunciation clarity in the generated speech, making it more intelligible to ASR systems. Notably, the improvements for Tamil (from 55.08 to 83.29) and Malay (from 61.25 to 76.40) in human-rated pronunciation accuracy are substantial, highlighting the effectiveness of the proposed approach in low-resource language settings. These results confirm the capability of \textit{MultiGen} to advance multilingual speech generation.

\begin{table}
\footnotesize
\centering
\caption{Comparison of speech intelligence results from subjective evaluation and objective evaluation by CER in Singaporean-accent Mandarin, Malay and Tamil.}
\vspace{-0.2cm}
\begin{tabular}{lrr}   
\toprule
\textbf{Singaporean-accent Mandarin} \\\midrule
Speech Intelligence & CosyBase & MultiGen \\\midrule
Objective Evaluation: CER (\%) & 10.70 & \textbf{4.00 }\\
Subjective Evaluation: Human & 74.37 &\textbf{ 83.40} \\\bottomrule
\textbf{Tamil} &  &  \\\midrule
Speech Intelligence & VITS & MultiGen \\\midrule
Objective Evaluation: CER (\%)& 9.60 & \textbf{2.10} \\
Subjective Evaluation: Human & 55.08 & \textbf{83.29 }\\\midrule
\textbf{Malay} &  &  \\\midrule
Speech Intelligence & VITS & MultiGen \\\midrule
Objective Evaluation: CER (\%)& 3.40 & \textbf{2.30 }\\
Subjective Evaluation: Human & 61.25 & \textbf{76.40} \\
\bottomrule
\end{tabular}
\label{cer}
\vspace{-0.5cm}
\end{table}

\section{Conclusion}
We propose a multilingual generator, \textit{MultiGen}, designed specifically for child-friendly speech generation. Our approach leverages advanced LLM architectures and culturally relevant, age-appropriate training strategies, and improve the text-to-speech performance in low-resource languages including Malay, Tamil, and Singaporean-accented Mandarin. The proposed approach can support a supportive and engaging communication environment with child-friendly voice, enhancing multilingual listening and speaking participation. Synthesized multilingual speech samples from different models are available for verification at the link \footnote{\url{https://xiaoxue1117.github.io/icmi2025demo/}}.

\section{Acknowledgment}
This research is supported by A*STAR under its Japan-Singapore Joint Call: Japan Science and Technology Agency (JST) and A*STAR 2024 (R24I6IR136), and by the National Research Foundation, Prime Minister’s Office, Singapore under its Campus for Research Excellence and Technological Enterprise (CREATE) programme (DesCartes). The educational use case is built on SingaKids \cite{liu2025singakids} from A*STAR.

\clearpage

\input{sample-sigconf.bbl}

\appendix

\end{document}